\documentclass[prl,twocolumn]{revtex4}

\usepackage{graphicx}

\begin{document}

\title{Reverse Schottky-Asymmetry Spin Current Detectors}

\author{Yuan Lu}
\altaffiliation{Present address: CNRS Institut Jean Lamour, Nancy, FR}
\author{Ian Appelbaum}
\altaffiliation{appelbaum@physics.umd.edu}
\affiliation{Center for Nanophysics and Advanced Materials and Department of Physics, University of Maryland, College Park MD 20742 USA}

\begin{abstract}
By reversing the Schottky barrier-height asymmetry in hot-electron semiconductor-metal-semiconductor ballistic spin filtering spin detectors, we have achieved: 1. Demonstration of $>$50\% spin polarization in silicon, resulting from elimination of the ferromagnet/silicon interface on the transport channel detector contact, and 2. Evidence of spin transport at temperatures as high as 260K, enabled by an increase of detector Schottky barrier height.
\end{abstract}

\maketitle

Measurement of spin transport in semiconductors has been achieved by detecting either spin-accumulation (density)\cite{LOU, SALIS, SHIRAISHIHANLE, JONKERLATERAL}, or spin current. The latter configuration has been achieved in silicon (Si) devices by exploiting the mean-free-path (mfp) asymmetry between spin-up and spin-down hot electrons in ferromagnetic (FM) thin films: electrons are ejected from the conduction band of the semiconductor forming the transport channel, and the ballistic component of this current is collected on the other side of the FM by a Schottky barrier.\cite{APPELBAUMNATURE}

\begin{figure}
\includegraphics[width=6.5cm, height=5cm]{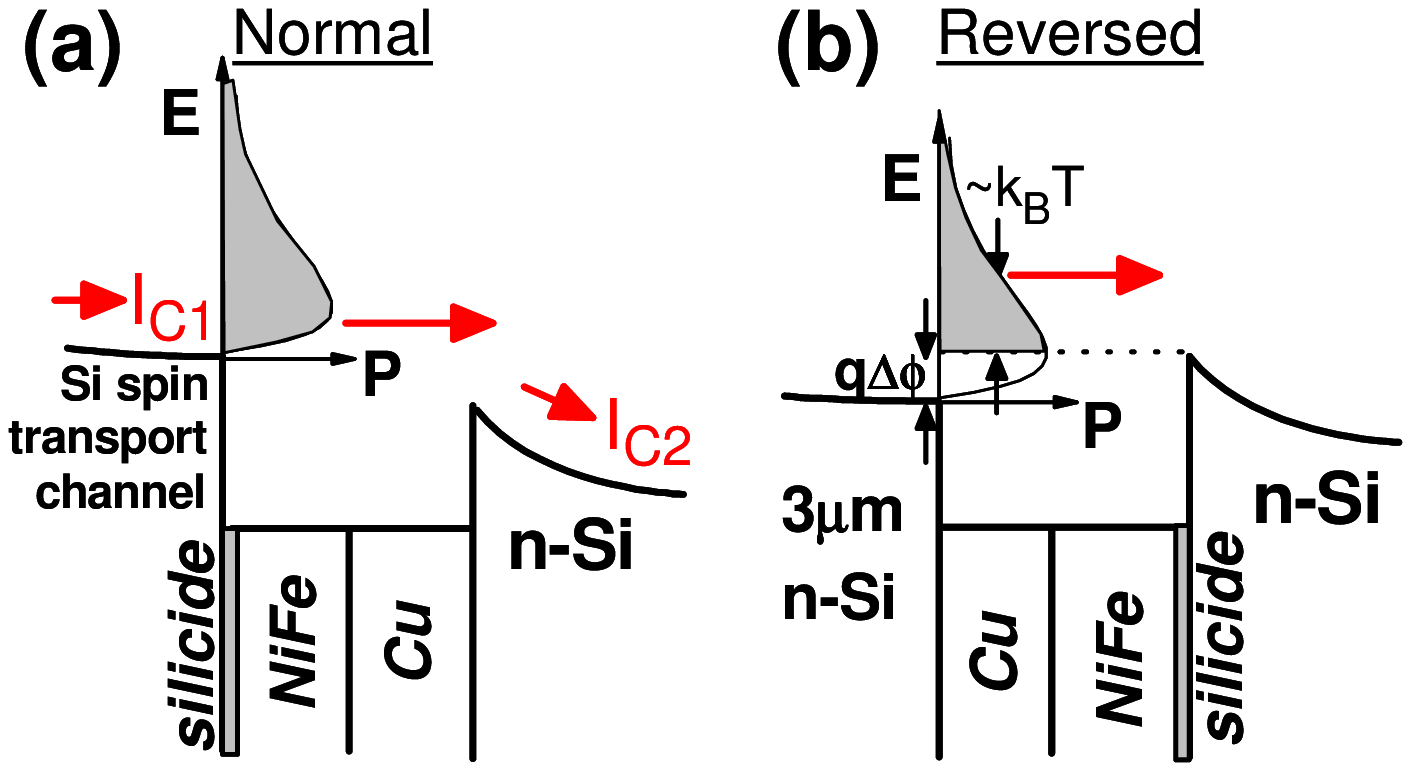}
\caption{\label{FIG1}
(a) Schematic band diagram of a semiconductor-metal-semiconductor hot electron spin detector with a lower detector-side Schottky barrier, used to increase collection efficiency in Refs. [\onlinecite{APPELBAUMNATURE, BIQINPRL, SPINFETEXPT, 2MM, 35PERCENT, LATERAL, DOPED, DEPHASING}] etc. (b) Spin detector with ``reversed'' Schottky barrier height asymmetry. Despite lowering the hot electron collection efficiency by preventing electrons in states close to the conduction band minimum from forming the signal current $I_{C2}$, this design eliminates transport through silicide before spin filtering. Resulting higher observed spin polarizations are shown in Fig. \ref{FIG2}). The higher collector Schottky barrier height enables higher temperature operation; see Fig. \ref{FIG4}).}
\end{figure}

The design of these detectors has incorporated the fact that to maximize the hot electron collection efficiency (namely, the ``transfer ratio'' of the ballistic current $I_{C2}$ to the current flowing through the channel $I_{C1}$), the collector Schottky barrier should be lower than the average hot electron energy defined by the Schottky barrier on the transport side. This design rule has resulted in the ``normal'' detector shown in Fig. \ref{FIG1}(a), where hot electrons ejected from Si on the left first pass through the NiFe ferromagnet and then over the (lower) Cu/n-Si Schottky barrier. Although this design has enabled many fundamental measurements (e.g. spin transport through the full wafer thickness\cite{BIQINPRL}, across 2 mm in quasilateral devices\cite{2MM}, in doped Si\cite{DOPED}, and long-distance lateral spin transport close to the buried Si/SiO$_2$ interface\cite{LATERAL}), the FM/Si interface present reduces the observed spin polarization. This is due to the presumed formation of a magnetically-dead silicide\cite{35PERCENT}, and the low Cu/n-Si Schottky barrier limits operation to $\approx$200K when thermionic leakage and its concomitant noise sufficiently dilutes the spin transport signal. 

In this Letter, we describe an alternative spin detector design where the FM is not in contact with the Si spin transport channel and the Schottky barrier height asymmetry is {\it reversed}. This design circumvents the problems discussed above and results in observations of both higher spin polarization and higher temperature operation.
  
\begin{figure}
\includegraphics[width=6.5cm, height=9.75cm]{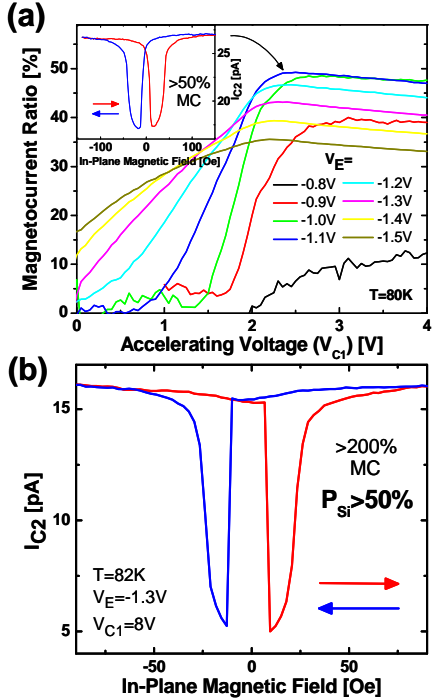}
\caption{\label{FIG2}
(a) Magnetocurrent ratio for MTJ-type injector devices (see Ref. [\onlinecite{35PERCENT}]) with 3$\mu$m-thick n-Si transport channel and the reverse asymmetry spin detector shown in Fig. \ref{FIG1}(b), as a function of accelerating voltage for several injection voltages $V_E$ at 80 K. Inset: A maximum magnetocurrent of over 50\% was measured for $V_E$=-1.1V and $V_{C1}$=2.4 V, as shown in the spin-valve measurement performed in an in-plane magnetic field. (b) If a ballistic spin filtering injector is used (see Ref. [\onlinecite{SPINFETEXPT}]), the injection and detection currents are smaller but spin polarization of $>$50\% can be observed. }
\end{figure}

Our ``reversed'' detector design is shown in Fig. \ref{FIG1} (b). Because the NiFe/n-Si Schottky barrier is higher than Cu/n-Si, electrons with low energy close to the conduction band edge on the transport side have no possibility to couple with conduction band states in the n-Si collector on the other side, and the detector hot electron collection efficiency transfer ratio is indeed lower. Spin-valve transistors\cite{SVT1, SVT2, SVTREVIEW} have been demonstrated in this reversed asymmetry as well.\cite{JANSENJMMM, JANSENJAP} However, the first benefit of this detector design is that by removing the possibility of silicide formation in the region before the electron spin is analyzed in the NiFe bulk, higher spin polarizations can be observed. This is because non-magnetic silicides of bulk elemental ferromagnet ions have large, randomly oriented magnetic moments that cause massive spin relaxation.\cite{SILICIDE1, SILICIDE2} Removal of a FM/Si interface at the injector side was previously shown to increase spin polarization from $\approx$1\%\cite{APPELBAUMNATURE} to $\approx$37\%.\cite{SPINFETEXPT}   Using a magnetic-tunnel-junction (MTJ)-type spin injector\cite{35PERCENT} and a 3$\mu m$-thick 1-10$\Omega\cdot cm$ n-type Si transport channel in a vertical device geometry\cite{DOPED}, we can observe a magnetocurrent change ($MC=(I_{C2}^P-I_{C2}^{AP})/I_{C2}^{AP}$, where the superscripts $P$ and $AP$ refer to parallel and antiparallel magnetic configuration, respectively) of greater than 50\% (shown with in-plane magnetic field measurements in Fig. \ref{FIG2}(a)). This is significantly higher than the 6\% previously reported in devices using identical injector and the ``normal'' detector design.\cite{DOPED} Because this injector suffers from greatly reduced injected polarization at high tunnel-injector bias ($V_E$), we also present measurements from devices with a ballistic-spin-filtering spin injector (similar to Ref. [\onlinecite{SPINFETEXPT}] where $\mathcal{P}_{Si}=$37\% spin polarization was reported). In this case (shown in Fig. \ref{FIG2}(b)), over 200\% magnetocurrent (corresponding to $\mathcal{P}_{Si}=\frac{MC}{MC+2}>$50\% spin polarization in the Si transport channel) can be seen.      

\begin{figure}
\includegraphics[width=6.5cm, height=9.75cm]{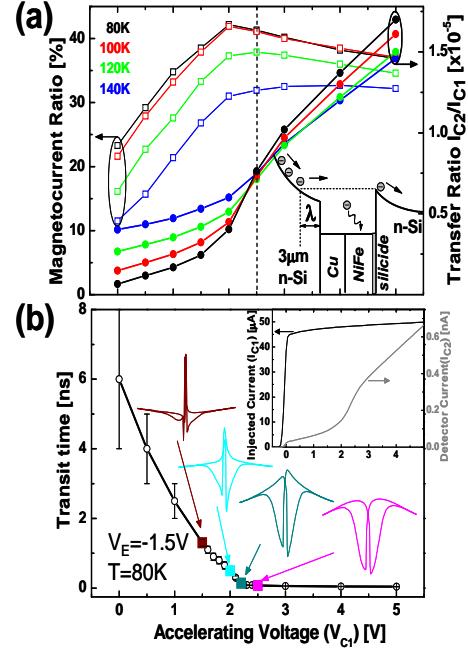}
\caption{\label{FIG3}
(a) The reverse-asymmetry detector magnetocurrent ratio (left axis) shows constant temperature dependence, whereas the transfer ratio $I_{C2}/I_{C1}$ (right axis) increases with temperature due to redistribution of electrons to higher energies $k_BT$ for low accelerating voltage drop $V_{C1}<$2.5 V across the n-Si transport channel. For higher accelerating voltage, the temperature trend reverses due to large kinetic energy imparted by the internal electric field over a mean-free-path $\lambda$, and effective elimination of the barrier asymmetry, as shown in the inset. (b) Hanle spin precession measurements at 80 K reveal the spin transit time, confirming that the crossover at $V_{C1}\approx$2.5 V corresponds to the elimination of a confining potential in the transport channel n-Si resulting from detector-side depletion. Inset: Increase in transfer ratio is due to $I_{C2}$. All data taken at MTJ-injector voltage of $V_E$=-1.5 V.}
\end{figure}

Because the low-energy portion of the electron distribution (determined by the product of Boltzmann occupation distribution and the conduction band density of states) does not play a role in the collected current $I_{C2}$, this detector has a lower transfer ratio $I_{C2}/I_{C1}$ than the normal detector design. However, as the thermal energy $k_BT$ increases, a greater fraction of electrons occupy the high-energy tail of the distribution and the transfer ratio should increase. This behavior is clearly shown at low accelerating voltages in Fig. \ref{FIG3}(a), where the data (right axis) is from MTJ-type injector devices. Despite a consistent trend with temperature in the magnetocurrent (left axis), this transfer ratio trend reverses at accelerating voltages above $\approx$2.5V, where large electric fields  $E\approx 10^4$ V/cm can accelerate electrons to kinetic energy $E \cdot \lambda$ on the order of the Schottky barrier asymmetry within a mfp $\lambda$. Under these circumstances, all electrons can potentially contribute to the spin signal $I_{C2}$ (i.e. energy conservation is no longer an obstacle to transport) and inelastic scattering in the metal film bulk (which increases with temperature) largely determines the temperature dependence of the transfer ratio. The inset to Fig. \ref{FIG3}(a) illustrates this high-field effect. 

This explanation based on internal electric-field is confirmed by measurements where the magnetic field is perpendicular to the plane, and Hanle spin precession and dephasing is detected. In Fig. \ref{FIG3}(b), transit times extracted from the width of the quasi-Lorentzian Hanle data measured in the range $\pm$850Oe show that at slightly lower accelerating voltages than the crossover value in the transfer ratio temperature trend (see vertical dotted line in Fig. \ref{FIG3}(a)), the transit time collapses. This corresponds to a removal of the confining potential in the conduction band of the transport channel caused by depletion regions on both the injector and detector sides. In a fully depleted model, the donor density consistent with this catastrophe at $V^{*}_{C1}\approx$2V is given by $n=\frac{2\epsilon V^{*}_{c1}}{qL^2}=$3$\times$10$^{14}$ cm$^{-3}$. This is somewhat less than in similar low-doped Si transport channel devices previously reported\cite{DOPED}, which can easily be due to doping variation between wafers.

For further confirmation of the low-field ($V_{C1}<2.5$ V) device behavior, we model the temperature dependence of the detector transfer ratio by calculating the fraction of electrons in the transport channel which have more energy than the Schottky asymmetry $q\Delta \phi$ via a semiclassical approximation (see Fig. \ref{FIG1}(b) for illustration): 

\begin{equation}
\frac{\int_{q\Delta \phi}^{\infty}e^{-E/k_BT}\sqrt{E}dE}{\int_{0}^{\infty}e^{-E/k_BT}\sqrt{E}dE},
\label{EQ1}
\end{equation}
 
\begin{figure}
\includegraphics[width=6.5cm, height=9.75cm]{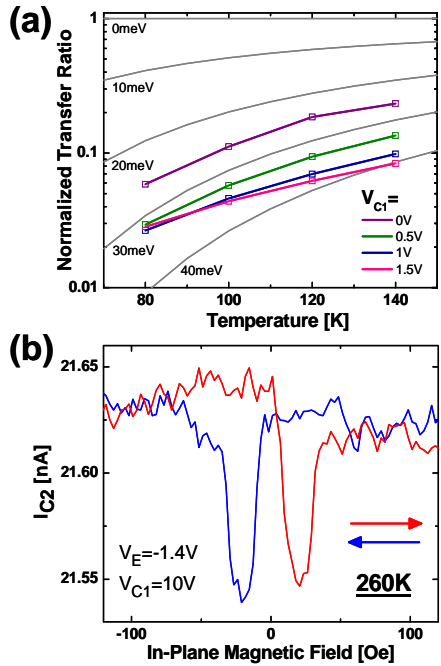}
\caption{  \label{FIG4}
(a) Comparison of simple model prediction of temperature dependence (Eq. \ref{EQ1}) to experimental results with the spin detector shown in Fig. \ref{FIG1}(b), showing that the transfer ratio increases with temperature, consistent with a $\approx$30 meV Schottky barrier asymmetry. (b) The 30 meV-higher Schottky barrier provided by the NiFe/n-Si collector in the ``reversed'' detector  enables operation up to temperatures of 260K, substantially higher than ``normal'' detectors ($\approx$200K).}
\end{figure}

\noindent where all other factors (effective mass, etc.) cancel, and we have not taken into account the role of inelastic scattering in the metal thin films. Note that this model also does not incorporate the high-field behavior schematically illustrated in the inset to Fig. \ref{FIG3}(a). Therefore, we compare the model prediction from evaluation of Eq. (\ref{EQ1}) in Fig. \ref{FIG4}(a) for several values of $q\Delta \phi$ to empirical transfer ratios only for $V_{C1}<$2.5 V. Furthermore, we normalize these empirical values by those from a similar device with ``normal'' detector (not shown) to remove the temperature dependence of the inelastic scattering in the metal thin films and reveal only the effect of Schottky asymmetry. As can be seen in Fig. \ref{FIG4}(a), our results are consistent with a Schottky asymmetry of $\approx$30meV, close to the known difference between Cu and Ni/Fe Schottky barriers on n-Si.\cite{SZEBOOK}

This modest 30meV increase in collector Schottky barrier actually has substantial impact on the operating temperature of these spin transport devices. Since Schottky leakage current is super-exponentially dependent on temperature, the high-temperature limit to operation is very sensitive to barrier height. With Cu/n-Si detectors, our temperature of operation is limited to $\approx$200K before thermionic leakage sufficiently degrades performance. In this alternative ``reversed'' Schottky asymmetry detector, we can now make spin-valve measurements consistent with spin transport up to 260K, as shown in Fig. \ref{FIG4}(b) using the 3$\mu$m-thick vertical n-Si devices with MTJ-type injectors previously described. Therefore, we expect the spin detector discussed here to have implications for studying true spin transport in semiconductors under conditions approaching room-temperature (300K).

This work was supported by the Office of Naval Research and the National Science Foundation. We acknowledge the support of the Maryland NanoCenter and its FabLab.

\end{document}